\documentclass[]{aa520}
\usepackage{graphicx}
\usepackage{natbib}
\bibpunct{(}{)}{;}{a}{}{,}
\begin{document}

\title{Modeling of Dust Scattering in the Coalsack}
\author{Jayant Murthy \and  P. Shalima}
\institute{The Indian Institute of Astrophysics, Koramangala, Bangalore 560 034}
\offprints{Jayant Murthy,
\email{jmurthy@yahoo.com}}
\date{Received / Accepted}
\abstract{\citet{Mcs} discovered intense UV (1100 \AA) emission from the direction of the Coalsack Nebula. We have used their results in conjunction with  a Monte Carlo model for the scattering in the region to show that
  the scattering is from dust in the foreground of the Coalsack molecular cloud. We have constrained the albedo of the grains to 0.4 $\pm$ 0.1. This is the first determination of the albedo of dust in the diffuse ISM in the FUV. 
}

 \maketitle

\section{Introduction}

Although a potentially useful test of models of interstellar dust grains, measurements of the optical
parameters --- the
albedo ($\it{a}$) and the phase function asymmetry factor ($\it{g}$) ---
of the grains have been too uncertain to have been of much utility \citep[for a recent review see][]{Dr03}.
 Both methods
used to investigate the scattering properties
have been problematic: reflection nebulae because an 
uncertain geometry can heavily
influence the derived parameters \citep{Ma02} ; and the diffuse background because of its faintness and 
because there is often
a tradeoff between $a$ and $g$ which allows neither to be tightly constrained \citep{Dr03}.

The Coalsack Nebula, one of the most prominent dark nebulae in the Southern Milky Way, offers an excellent
location
for the determination of the scattering function of the diffuse grains, particularly in the UV where 
 \citet{Mcs}
found
 it to be one of the brightest sources of diffuse emission in the sky. 
Without detailed modeling,
 they were unable to provide useful contraints on the optical constants of the grains but did 
suggest that most
of the observed emission was due to forward scattering of photons from three of the brightest UV stars
in the sky (Table \ref{star_inf})
by foreground dust, rather than back-scattering from dust in the molecular cloud.  We note that \citet{Mat70}
observed scattering from the Coalsack in the visible which he interpreted as being
due to scattering from the Coalsack.

In this paper, we have reinterpreted the $\it{Voyager}$ observations of \citeauthor{Mcs}
using improved distances for the three
stars, a detailed model for the interstellar dust distribution and a Monte Carlo model for the grain scattering. 
In agreement with \citeauthor{Mcs}, the observed radiation is  dominated by scattering from dust in the foreground cloud, rather than the Coalsack molecular cloud. 
The albedo is tightly constrained to a value of 0.4 $\pm$ 0.1 at 1100 \AA. 
\begin{table}
\caption[]{Properties of stars in our model}
\label{star_inf}
\begin{tabular}{r c c c c}
\hline
\hline
Name&$\it{l}$&$\it{b}$&distance&flux(1100 \AA )\\
    & (deg) & (deg) & (pc) & (photons s$^{-1}$ \AA$^{-1}$ ) \\
\hline
$\alpha$ Cru &300.13&-0.36&98.3&8.86$\times$10$^{8}$\\
$\beta$ Cru &302.46&3.18&108.1&9.70$\times$10$^{8}$\\
$\beta$ Cen &311.77&1.25&161.3&2.53$\times$10$^{9}$\\
\hline
\end{tabular}
\end{table}

\section{Model}

We have developed a generalized Monte Carlo model to simulate the scattered emission from a star in an 
arbitrary scattering
geometry. Each photon from the star is emitted in a random direction and continues in that direction 
until an interaction occurs, the probability of which depends on the local dust density. 
At the point of interaction, we reduce the photon's effective weight by a factor
of $\it{a}$, the grain albedo, and calculate a new direction using the
Henyey-Greenstein \citep{H.G} scattering function:
\begin{equation}
\phi(\theta)=\frac{(1-{\it g}^2)}{(1+{\it g}^2-2{\it g}cos(\theta))^{3/2}}
\end{equation}
In Eqn. 1, $\it{g}$ is the phase function asymmetry factor (defined as $\langle cos(\theta)\rangle $) 
and $\theta$ is the angle of scattering. If {\it g} is close to zero, the scattering is 
nearly isotropic while a value of {\it g} near 1 implies strongly forward scattering grains.

\begin{figure}[t]
\centering
\includegraphics[angle=-90,width=9cm]{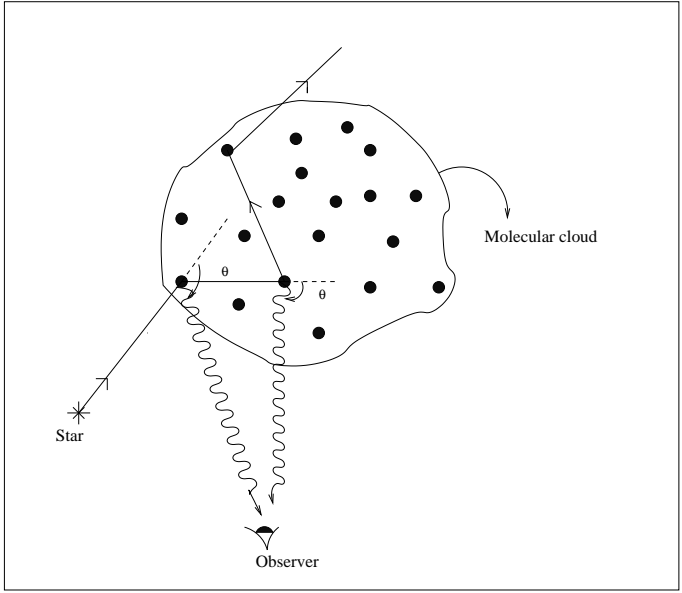}
\caption{In our Monte Carlo model, photons are emitted by the star in a random direction and proceed until an interaction occurs. After each interaction, each photon is reemitted in a new direction as determined by the scattering function. In order to save computational time, a part of the energy at each interaction is redirected to the observer.}   
\label{schematic}%
\end{figure}

We follow the photon through a sequence of interactions until it either leaves the area we are 
considering or its 
intensity drops to a negligible value. To save computational time, a part of every photon was redirected
to the observer at each interaction. This led to a convergence of the model solution in a few million
iterations, after which the results were scaled to the stellar output.
A schematic of our model is shown in Fig. \ref{schematic}.

Fortunately for us, only three early-type stars (Table \ref{star_inf}) dominate the FUV radiation field 
in the Coalsack. The nebula, itself, blocks any light from more distant stars and the other foreground stars 
are all cool stars with negligible FUV emission. We have used data from the Hipparcos catalog \citep{Hipparcos}
to specify the stellar spectral types, locations, and distances. The flux from each star was calculated using a \citet{Kurucz} model scaled to the observed IUE flux at 1500 \AA.
The total number of photons emitted by each star
could then be directly calculated.

\begin{figure}[t]
\centering
\includegraphics[angle=90,width=9cm]{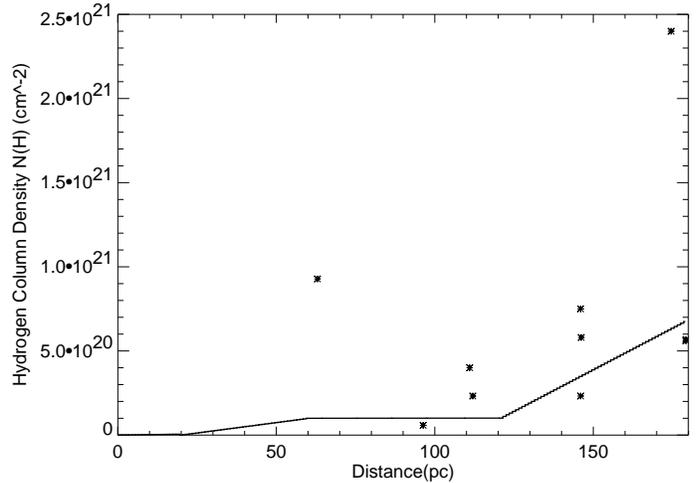}
\caption{The hydrogen column density (N(H)) is plotted for a number of stars (asterisks) in the central
$2\degr \times 2\degr$ of the field. Based on these measurements, we have distributed the dust as shown by
the solid line. We use similar plots in other regions to constrain the dust over the entire 8$\degr \times 
8\degr$ field.}
\label{dust_dist}
\end{figure}

The dust distribution, as usual, is more difficult to characterize. The molecular cloud comprising
the Coalsack is clearly delimited by the CO contours of \citet{Dame} which we have converted into
a total hydrogen column density using the $\it{N}$(H$_{2}$)/$\it{W}_{CO}$ ratio found by \citet{Blh2}. We have arbitrarily assumed that the cloud is 1 pc 
thick, (defined by our bin size) which gives local space densities of between 200 and 1000 cm$^{-3}$ 
in the cloud, well within the canonical range for molecular clouds \citep{Spitzer}. 
\begin{table}
\label{dust_dens}
\end{table}
The distance of the Coalsack is between 180 and 200 pc
from the Sun \citep{Frcs}, behind any of the stars of Table \ref{star_inf}. 
There is virtually no interstellar matter in this direction upto a distance of about 40 pc, except for the Local Cloud, which only has a column density of about 5$\times$10$^{18}cm^{-2}$ \citep[see][]{Fr03}. Beyond 40 pc, we have used published data from a variety of sources to constrain the dust distribution. In practice, we found that there were sufficient stars to determine the dust distribution in 2$\degr \times 2\degr$ squares, one example of which is shown in Fig. \ref{dust_dist}.

\setcounter{table}{2}
\begin{table}
\caption[]{Observed fluxes in the Coalsack}
\label{voy_res}
\begin{tabular}{r c c c c}
\hline
\hline
$\it{l}$ & $\it{b}$ & Intensity &Ref. \\
(deg) & (deg) &(observed)\footnotemark[2]\\
\hline
303.7 & 0.8 & 12300 $\pm$ 800\footnotemark[3]&\citet{Mcs}\\
303.7 & 0.8 & 15500 $\pm$ 800\footnotemark[3]&\citet{Mcs}\\
305.2 & -5.7 & 8300 $\pm$ 1200\footnotemark[3]&\citet{Mcs}\\
304.6 & -0.4 & 12300 $\pm$ 1200\footnotemark[3]&\citet{Mcs}\\
301.7 & -1.7 & 18900 $\pm$ 400&\citet{Mu98}\\\hline
\end{tabular}

\thanks{\footnotemark[2] photons cm$^{-2}$ s$^{-1}$ sr$^{-1}$ \AA$^{-1}$.}\\       
\thanks{\footnotemark[3] Fluxes have been reduced due to an incorrect calibration.}
\end{table}
We have run our model for various combinations of the optical constants and finally compared with
the $\it{Voyager}$ results of \citet{Mcs} in Table \ref{voy_res}. The main sources of uncertainty
in the modeling lie in the dust distribution, which may be poorly characterized, and in the use of the 
Henyey-Greenstein scattering function, which has been found to poorly represent the UV scattering of 
radiation by \citet{Draine}. We have empirically accounted for these uncertainties by simply increasing 
the error bars associated with the data, which are much smaller than the model uncertainties, 
such that the minimum $\chi^2$ $\equiv$ 1.

\section {Results and Discussion}
\begin{figure}[t]
\centering
\includegraphics[angle=+90, width=9cm]{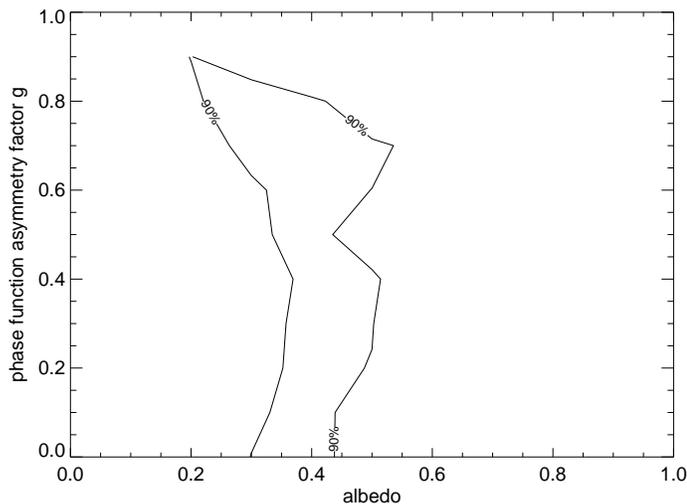}
\caption{A 90\% confidence contour is plotted for $\it{g}$ versus $\it{a}$. Although we can place few constraints
on $\it{g}$, we can constrain $\it{a}$ to lie between about 0.3 and 0.5.}
\label{agcntr}
\end{figure}

\begin{figure}[ht]
\centering
\includegraphics[width=9cm]{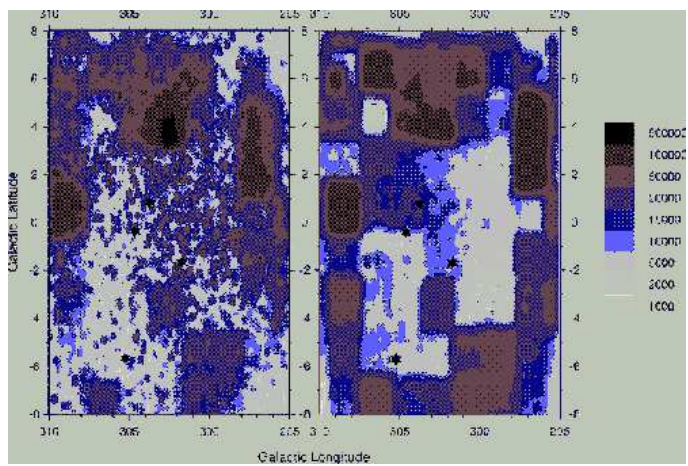}
\caption{The emission predicted by our model with different values of $\it{a}$ and $\it{g}$ are shown here
- $\it{a}$ = 0.3; $\it{g}$ = 0.9 on the left and $\it{a}$ = 0.5; $\it{g}$ = 0.1 on the right
and the scale is shown on the right. The locations of the 5 $\it{Voyager}$ observations are shown 
in the images as stars - note that one of the locations was observed with 
 both $\it{Voyager} 1$ and $\it{Voyager} 2$.}
\label{scalight}
\end{figure}

By comparing our model runs with the observations (Table \ref{voy_res}), we 
have found that the best fit value for the albedo is $\it{a}$ = 0.4 $\pm$ 0.1  
with 90\% confidence contours as shown in Fig. \ref{agcntr}.
 Unfortunately, we cannot similarly constrain $g$. The primary reason for this is seen in 
Fig. \ref{scalight} where we have plotted the distribution of the scattered starlight for both isotropic
and forward scattering grains. 
Because the locations of the 5 $\it{Voyager}$ targets (shown as large stars in Fig.
\ref{scalight}) were essentially chosen at random, the observations at those positions can be matched
by any value of $g$.

There are very few determinations of the optical constants of interstellar grains in the FUV and none have been of grains in the diffuse ISM  \citep[see][]{Dr03,Gord04}. The albedo derived here is consistent with that from observations of reflection nebulae \citep{Witt93,Cal94,Burgh02}.
Although our observations cannot distinguish between different values of $g$, we have
proposed new observations in regions selected such that we can unambiguously determine the optical constants. 

\begin{acknowledgements}
This research has made use of the SIMBAD database, operated at CDS, Strasbourg, France and NASA's Astrophysics Data System Bibliographic Services. 
\end{acknowledgements}

\bibliography{clsack}

\setcounter{table}{1}
\begin{table*}
\begin{center}
\caption[]{Table of column density measurements for various stars in the region.

References: 1. SIMBAD; 2. \citet{Frcs}; 3. \citet{Seidensticker}; 4.\citet{Franco}; 5. \citet{York}; 6. \citet{Shull}; 7. \citet{Bohlin}; 8. \citet{Craw91}}
\label{dust_dens}
\begin{tabular}{r c c c c c}
\hline
\hline
HD Number & Dist & N(H) & $\it{l}$ & $\it{b}$ & Ref.\\
          &(pc)  &($\times 10^{20}$$cm^{-2}$)&(deg)&(deg)& \\\hline
58806\footnotemark[2] & 294.9 &  20.8& 296.764 &  3.55&1\\
94493  & 980.4& 13.3& 289.0 & -1.20&1\\
97617  & 284.1&  8.7 & 293.5 & -6.04&1\\
98195  & 284.1&  5.8 & 294.9 & -8.53&1\\
99149  & 157.7&  5.2& 294.1 & -4.47&1\\
99857  & 483.1& 23.7& 294.8 & -4.90&1\\
100101 & 284.9&  45.2& 294.5 & -3.48&1\\
100666 & 178.3&  11.6& 294.6 & -2.33&1\\
100990 & 255.1&  11.0 & 296.4 & -7.33&1\\
100927 & 238.7&  42.3& 293.5 &  2.06&1\\
101190 & 237.1& 24.3& 294.8 & -1.50&1\\
101903 & 249.3&  41.7& 295.9 & -3.44&1\\
101929 & 145.6& 9.3& 293.3 &  6.40&1\\
101190 & 237.1&  10.9& 294.8 & -1.49&1\\
101966 & 135.8&  5.2 & 296.8 & -6.42&1\\
102461 & 251.8&  30.9& 294.4 &  4.11&1\\
102544 & 255.1&  7.5& 297.0&-5.63&1\\
102349 & 709.0&  30.1& 295.0 &  1.63&1\\
102728 & 190.6&  53.3& 295.4 &  1.52&1\\
102386 & 189.4&  30.7& 295.5 & -0.36&1\\
103066 & 303.0&  12.7& 296.6 & -2.46&1\\
103168 & 362.0&  44.0 & 296.2 & -0.35&1\\
103079 & 103.7&  0.2& 296.7 & -3.05&8\\
103101 & 104.5&  1.0   & 294.9 &  4.95&1\\
104432 & 628.0&  6.4& 297.2 & -0.29&1\\
104841 & 230.9&  8.5 & 297.6 & -0.77&1\\
104936 & 251.8&  34.2& 298.9 & -7.49&1\\
104479 & 216.4&  26.1& 298.5 & -6.75&1\\
104125 &  96.7&  5.8 & 295.9 &  4.99&1\\
104564 & 253.8&  2.3& 296.2 &  5.67&1\\
105017 & 395.0&  33.0& 298.1 & -2.44&1\\
105194 & 254.5&  4.0& 297.6 &  1.21&1\\
105822 & 288.2&  33.0 & 299.2 & -5.68&1\\
105907 & 460.8 &  27.8& 297.5 &  5.65&1\\
106521 & 411.0 &  11.6& 298.6 &  1.53&1\\
106490 & 111.6 &  1.1& 298.2 &  3.79&7\\
107652 & 313.5&  4.9& 299.6 &  0.47&1\\\hline
\end{tabular}
\end{center}
\thanks{\footnotemark[2] Hipparcos number.}
\end{table*}

\setcounter{table}{1}
\begin{table*}
\centering
\caption{continued}
\label{dust_dens}
\begin{tabular}{r c c c c c}
\hline
\hline
HD Number & Dist & N(H) & $\it{l}$ & $\it{b}$ & Ref.\\
          &(pc)  &($\times 10^{20}$$cm^{-2}$)&(deg)&(deg)& \\\hline
107821 & 97.9& 5.8& 299.9 & -1.16&1\\
107411 & 146.0 &  7.5 & 302.1  & -1.34&4\\
107978 & 295.9& 6.6& 299.7 &  1.82&1\\
107082 & 184.5&  34.2& 298.9 &  2.41&1\\
107983 & 191.2&  10.4& 300.7 & -7.96&1\\
107789 & 151.1&  2.3& 299.1 &  5.88&1\\
108395 & 268.1&  34.8& 299.8 &  4.40&1\\
108750 & 133.9& 1.2& 300.4 &  1.31&1\\
108804 & 217.4& 1.2& 300.5 &  0.95&1\\
108813 & 68.2& 1.7& 300.5 &  1.37&1\\
108531 & 343.6& 8.1& 300.2 &  0.68&1\\
108610 & 378.8& 8.4& 300.3 &  0.88&1\\
108939 & 2272.7& 6.9& 300.5 &  1.89&1\\
108447 & 166.6 &  26.1& 301.0 & -7.76&1\\
108248 & 98.3&  0.4& 300.1 & -0.36&5\\
108355 & 201.6&  8.4 &  300.3 & -1.04&4\\
108483 & 135.8& 3.6& 299.1 &  12.50&1\\
108671 & 186.5& 2.9& 300.8 & -3.82&1\\
108608 & 1068.3& 5.5& 300.2 &  1.82&1\\
108999 & 636.6& 6.6& 300.6 &  0.96&1\\
109000 & 73.7&  0.3& 300.8 & -0.72&1\\
109266 & 589.8& 5.8& 300.9 &  1.03&1\\
109493 & 660.8& 8.9& 301.1 & -0.06&1\\
109504 & 676.1& 11.0& 301.1 &  1.31&1\\
109810 & 1015.8& 9.8& 301.4 &  1.03&1\\
109152 & 110.3&  8.1& 301.3 & -6.02&1\\
109475 & 83.6& 3.8& 301.1 &  1.07&1\\
109550 & 403.2& 7.5& 301.2 & -0.02&1\\
109801 & 105.5&  3.7 & 301.6 & -2.99&3\\
109165 & 156.3&  8.1& 301.3 & -6.02&1\\
109047 & 239.2& 13.3& 300.7 &  0.29&1\\
109199 & 387.6& 14.2& 301.1 & -3.31&1\\
109478 & 125.3& 4.1& 301.3 & -2.46&1\\
109563 & 188.3& 2.4& 301.5 & -4.34&1\\
109614 & 313.5& 4.1& 301.5 & -4.38&1\\
109891 & 151.3& 13.3& 301.5 &  0.31&1\\
109993 & 277.0& 11.6& 301.8 & -4.19&1\\
109777 & 121.4&  6.9& 301.1 &  4.95&1\\
109993 & 277.0 &  10.4& 301.8 & -4.19&1\\
113153 & 120.9& 1.2&  304.0 &  -5.19&1\\
114911 & 124.4&  0.6   & 305.2 & -5.12&1\\\hline
\end{tabular}
\end{table*}

\setcounter{table}{1}
\begin{table*}
\centering
\caption{continued}
\label{dust_dens}
\begin{tabular}{r c c c c c}
\hline
\hline
HD Number & Dist & N(H) & $\it{l}$ & $\it{b}$ & Ref.\\
          &(pc)  &($\times 10^{20}$$cm^{-2}$)&(deg)&(deg)& \\\hline
115267 & 231.5&  11.6& 305.5  & -4.21&1\\
112254 & 255.1&  4.6 & 303.4  & -4.75&1\\
112938 & 189.3&  8.1 & 303.9  & -5.0&1\\
115583 & 169.5&  2.8 & 305.7 & -4.64&1\\
113558 & 178.9&  2.9 & 304.3  & -5.73&1\\
116424 & 172.7&  20.2& 306.1 & -5.86&1\\
115286 & 225.2&  6.3 & 305.4 & -5.74&1\\
113590 & 141.8&  1.2 & 304.3 & -5.32&1\\
114142 & 144.3&  0.6   & 304.7 & -4.74&1\\
113607 & 540.5&  1.7 & 304.3 & -4.91&1\\
116230 & 163.4&  0.6   & 306.1  & -4.18&1\\
112764 & 73.9&  9.3& 304.1 &  6.94&1\\
115842 & 102.1&  30.7& 307.1&  6.83&1\\
115436 & 278.5&  8.9& 306.6 &  5.38&1\\
111302 & 190.5&  11.6& 302.6 &  4.42&1\\
113455 & 184.2&  15.0 & 304.7 &  4.24&1\\
114808 & 165.5&  26.6& 305.9 &  4.42&1\\
117171 & 188.6&  34.2& 308.1 &  4.92&1\\
110390 & 105.1&  8.1 & 301.8 &  1.83&1\\
110772 & 223.7&  30.1& 302.1 &  2.78&1\\
113199 & 139.2&  8.7 & 304.4 &  2.84&1\\
112556 & 193.7&  18.6& 303.8 &  4.38&1\\
116087 & 108.7&  0.4 & 306.7 &  1.65&8\\
116780 & 202.4&  9.3& 307.4 &  2.85&1\\
119385 & 220.3&  12.7& 309.6 &  2.30&1\\
113919 & 306.7&  3.5& 304.6 & -4.97&1\\
116749 & 138.3&  1.0   & 306.5 & -4.12&1\\
116849 & 480.0&  20.8& 306.6 & -3.67&1\\
117651 & 106.9&  1.0   & 307.3 & -3.11&1\\
117652 & 221.7&  29.5& 307.1 & -4.20&1\\
113956 & 287.0&  8.1& 304.5 & -6.90&1\\
115286 & 225.2&  6.3& 305.4 & -5.74&1\\
116865 & 152.9&  11.0 & 306.4 & -5.24&1\\
117445 & 125.6&  8.7 & 306.6 & -6.65&1\\
116106 & 178.9&  10.4& 305.8 & -6.10&1\\
118229 & 261.1&  12.1& 307.1 & -6.47&1\\
110511 & 192.7&  8.1& 302.3 & -8.17&1\\
111442 & 253.1 &  9.3& 302.9 & -8.75&1\\
114887 & 289.8 &  12.6& 304.9 & -7.69&1\\
118684 & 293.2 &  10.4& 307.2 & -7.52&1\\\hline
\end{tabular}
\end{table*}

\setcounter{table}{1}
\begin{table*}
\centering
\caption{continued}
\label{dust_dens}
\begin{tabular}{r c c c c c}
\hline
\hline
HD Number & Dist & N(H) & $\it{l}$ & $\it{b}$ & Ref.\\
          &(pc)  &($\times 10^{20}$$cm^{-2}$)&(deg)&(deg)& \\\hline
118344 & 134.4 &  26.7& 306.9 & -7.93&1\\
118522 & 308.6&  27.1& 306.9 & -8.29&1\\
118846 & 205.1&  38.4& 308.6 & -0.17&1\\
119661 & 510.0&  12.1& 308.8 & -2.43&1\\
111037 & 117.7&  0.9 & 302.4 &  1.64&2\\
112123 & 179.2&  5.6 & 303.4  &  0.30&3\\
110640 & 194.1&  16.0 & 302.1 &  1.33&3\\
111580 & 111.0&  4.0   & 302.9 & -2.04&3\\
110151 & 751.9& 11.3& 301.6 &  1.93&1\\
110020 & 108.3& 3.6& 301.8 & -3.67&1\\
111283 & 1250.0& 11.7& 302.7 & -2.72&1\\
112607 & 304.9& 15.6& 303.8 & -0.78&1\\
112766 & 303.0& 4.4& 303.8 & -3.99&1\\
113991 & 671.1& 20.5& 305.0 &  1.87&1\\
114603 & 680.3& 10.7& 305.5 &  1.30&1\\
110310 & 135.7& 4.9& 302.0 & -1.88&1\\
110477 & 78.8& 17.4& 301.9 &  1.71&1\\
110610 & 146.2& 5.8& 302.1 & -1.35&1\\
111161 & 97.7& 5.2& 302.6 & -4.26&1\\
111303 & 354.6& 9.2& 302.7 &  1.80&1\\
112109 & 96.3& 0.6& 303.3 & -0.77&1\\
112703 & 63.0& 9.3& 303.8 & -1.51&1\\
110062 & 507.6& 17.6& 301.7 & -0.61&1\\
110079 & 271.0& 11.3& 301.8 & -3.01&1\\
110163 & 467.3& 15.3& 301.7 &  0.86&1\\
110245 & 671.1& 6.4& 302.0 & -4.12&1\\
110715 & 352.1& 21.1& 302.2 & -2.10&1\\
110737 & 174.5& 24.0& 302.3 & -2.46&1\\
110830 & 163.4& 12.7& 302.3 &  0.67&1\\
110925 & 221.2& 10.4& 302.3 &  2.03&1\\
110946 & 925.9& 26.3& 302.4 & -2.05&1\\
111409 & 383.1& 15.3& 302.8 & -1.74&1\\
111992 & 518.1& 27.8& 303.2 & -0.30&1\\
112169 & 145.9& 2.3& 303.4 & -0.26&1\\
112295 & 320.5& 38.0& 303.6 &  1.53&1\\
112954 & 613.5& 28.7& 304.1 & -0.08&1\\
113191 & 265.3& 8.7& 304.2 & -1.94&1\\
113457 & 95.1& 1.5& 304.4 & -1.61&1\\
113689 & 232.5& 6.4& 304.6 & -1.74&1\\
114792 & 2777.0& 29.0& 305.5 &  0.10&1\\\hline
\end{tabular}
\end{table*}

\setcounter{table}{1}
\begin{table*}
\centering
\caption{continued}
\label{dust_dens}
\begin{tabular}{r c c c c c}
\hline
\hline
HD Number & Dist & N(H) & $\it{l}$ & $\it{b}$ & Ref.\\
          &(pc)  &($\times 10^{20}$$cm^{-2}$)&(deg)&(deg)& \\\hline
114012 & 540.5& 26.6& 304.9 & -0.40&1\\
114670 & 2702.7& 5.8& 305.4 & -0.53&1\\
114738 & 211.4& 8.7& 305.4 & -1.55&1\\
114739 & 236.4& 4.8& 305.3 & -2.27&1\\
111557 & 205.5& 3.5& 302.9 & -0.12&1\\
112045 & 196.0& 30.8& 303.3 &  1.07&1\\
112225 & 139.8& 1.7& 303.5 &  1.95&1\\
113348 & 337.2& 9.8& 304.4 &  1.37&1\\
110373 & 1447.0& 16.2& 301.9 & -0.12&1\\
110956 & 121.4& 3.6& 302.2 &  6.40 &1\\
112026 & 2476.7& 15.9& 303.3 &  1.98&1\\
112536 & 492.8& 17.1& 303.8 &  1.84&1\\
110433 & 879.0& 24.9& 302.0 & -0.33&1\\
110449 & 1488.5& 19.1& 301.9 &  1.88&1\\
110736 & 771.6& 20.3& 302.2 & -0.11&1\\
110975 & 616.9& 22.0& 302.4 & -0.28&1\\
111024 & 566.7& 22.0& 302.4 & -0.22&1\\
111687 & 618.2& 5.2& 303.0 &  1.79&1\\
111827 & 473.4& 16.8& 303.1 & -1.95&1\\
112637 & 978.2& 30.7& 303.8 & -0.45&1\\
113014 & 1612.5& 30.1& 304.1 &  0.67&1\\
113511 & 2721.2& 42.9& 304.5 & -1.22&1\\
113658 & 707.4& 11.3& 304.7 &  1.68&1\\
113968 & 1094.9& 13.9& 305.0  &  1.56&1\\
114317 & 1105.6& 7.8& 305.3 &  1.82&1\\
114318 & 681.2& 10.7& 305.3 &  1.40&1\\
114460 & 541.4& 5.8& 305.3 &  0.48&1\\
112485 & 2286.7& 16.2& 303.7 &  2.05&1\\
112078 & 110.4& 10.0& 303.4&     3.7&6\\
118716 & 115.2& 5.7& 310.2 &  8.70&1\\
120359 & 262.4 &  5.8 & 310.6 &  3.48&1\\
120768 & 83.6  &  4.6& 310.6 &  1.93&1\\
120891 & 235.8 &  7.5& 309.1 & -4.78&1\\
121796 & 195.3&  4.1& 310.0 & -3.60&1\\
122036 & 315.4 &  4.6& 309.2 & -7.03&1\\
122098 & 361.0 &  4.8& 310.5 & -2.39&1\\
122144 & 161.0&  0.4& 310.9 & -1.17&6\\
122451 & 161.0& 2.6& 311.8 &  1.25&1\\
124316 & 353.3 &  5.8 & 310.6 & -6.92&1\\\hline
\end{tabular}
\end{table*}

\end{document}